\documentstyle[preprint,aps,prd]{revtex}
\tightenlines
\newcommand{\be}{\begin{equation}}
\newcommand{\br}{\begin{array}}
\newcommand{\er}{\end{array}}
\newcommand{\beq}{\begin{equation}}
\newcommand{\ee}{\end{equation}}
\newcommand{\eeq}{\end{equation}}

\newcommand{\N}{{\cal N}}

\newcommand{\p}{\partial}

\def\ba{\begin{eqnarray}}
\def\ea{\end{eqnarray}}

\begin{document}

\preprint{hep-th/0004142}
\title {The M-theory dual of a 3 dimensional theory with reduced supersymmetry}
\author{Iosif Bena}
\address{University of California, Santa Barbara, CA 93106 \\ iosif@physics.ucsb.edu}
\date{\today}
\maketitle

\begin{abstract}
In a recent paper, Polchinski and Strassler found a string theory dual of a gauge theory with reduced supersymmetry. Motivated by their approach, we perturb the $\N=8$ theory living on a set of $N$ M2 branes to $\N=2$, by adding fermion mass terms. We obtain M-theory duals corresponding to M2 branes polarized into M5 branes, in  $AdS_4 \times S_7$. 
In the course of doing this we come across an interesting feature of the M5 brane action, which we comment on.
Depending on the fermion masses we obtain discrete or continuous vacua for our theories. We also obtain dual descriptions for domain walls, instantons and condensates.

\end{abstract}


\section{Introduction}

In the framework of the AdS-CFT duality \cite{ads} and of brane polarization \cite{myers}, Polchinski and Strassler \cite{joe} found a supergravity dual of a confining gauge theory by perturbing $AdS_5 \times  S_5$ with a 3 form field background. The AdS-CFT duality allowed them to extract information about a 4 dimensional $\N=1$ gauge theory. In particular they found a mapping between the gauge theory vacua and states corresponding to the D3 branes being polarized into NS5 and D5 branes.

We can apply the same philosophy in order to obtain the supergravity dual of a theory coming from perturbing the 3 dimensional $\N=8$ theory living on $N$ M2 branes. The approach is similar, but at some points subtle differences between string theory and M theory come to play a role.

As discussed in \cite{seiberg}, this 3 dimensional theory is obtained in the IR (strongly coupled) limit of a 3 dimensional $\N=8$ SYM. We know that the strongly coupled theory has 8 scalars and 8 Majorana fermions. Of these, 6 scalars and 6 fermions already can be paired into 3 hypermultiplets in the UV (the D2 brane theory). In the strongly coupled limit, SO(8) symmetry is restored, so the other scalar pairs up with the dualized $A_{\mu}$ and the 2 other fermions into a hypermultiplet. We can give masses to these 4 hypermultiplets (which appear as fermion masses in the Lagrangian), preserving $\N=2$ supersymmetry.

The fermions transform in the {\bf 8}$'$ of the SO(8) R-symmetry group. Thus, a fermion mass transforms in the ${\bf 35}_-$ of SO(8). By the AdS-CFT duality, giving mass to the fermions corresponds to turning on a nonnormalizable mode of the antiselfdual 4-form field strength in the 8 dimensional transverse space of the M2 branes.

In \cite{myers} it was observed that in a background p+3 form field, Dp branes become ``polarized''. The polarization was understood in 2 ways. In one picture the configuration with the D-branes spread on an $S^2$ with a p+2 brane charge was energetically favored to the configuration with the D-branes in the center in a background p+3 form field. In the other picture, the nonabelian scalars describing the position of the brane become noncommutative, which resulted in a p+2 brane charge. This was worked out in more detail in \cite{joe}. In the 3 brane case, for example, noncommutative configurations of the 3 chiral multiplets describe the polarization into a 3+2=5 brane.

Based on the above, we expect M2 branes to polarized also, when placed in a field configuration which couples with a higher brane (which can only be the M5 brane). We can understand M2 brane polarization easily in the first way - an M5 brane of geometry $R^3 \times S^3$ with M2 brane charge will have a supersymmetric minimum at a nonzero radius. This is what most of this paper will be on. 

Unfortunately the degrees of freedom of the M2 brane are not known, so the second picture is elusive. The weakly coupled theory is irrelevant. It describes the polarization of D2 branes into D4 branes, which is a different subject to be treated on its own \cite{d2}.

We can present at most a speculation of this type: for D3 branes, when we give mass to the 3 chiral multiplets the vevs become noncommutative. This can be interpreted as polarization into a 2-dimensional higher object. We expect on intuitive grounds that when given mass, 4 hypers becoming ``noncommutative'' (whatever that means if they are not matrices) represent somehow the polarization of a M2 brane into a 3 dimensional higher object. We will present a bit of support for this picture in chapter V.

\section {Perturbations of $ AdS_4 \times S^7$}

As it is by now standard lore, in the framework of the AdS-CFT duality, to each local CFT operator of dimension $\Delta$ correspond one normalizable and one nonnormalizable solution of the supergravity field equation. The coefficient of the nonnormalizable solution corresponds to the coefficient of the operator in the Hamiltonian, while the coefficient of the normalizable one corresponds to the vev of this operator \cite{vijay}.

According to the AdS-CFT conjecture the conformal field theory living on a $N$ M2 branes in dual to M theory living in the geometry they create, for very large $N$. 
This geometry is
$$ds^2=Z^{-2/3}\eta_{\mu \nu}dx^{\mu}dx^{\nu} + Z^{1/3}dx^idx^i$$
$$C^0_3= -{1 \over Z} dx^0 \wedge  dx^1 \wedge  dx^2, \ \ \ \ F^0_4 = d C^0_3,
\eqno(1)$$
where $\mu,\nu = 0,1,2$, $i,j = 3,...,10$.  For the case when the branes are coincident, the geometry becomes $AdS_4 \times S^7$, and:
$$Z = {R^6 \over r^6}, \ \ \ \ \  R^6 = 32 \pi^2 N M_{11}^{-6}, \eqno(2)$$
where $M_{11}$ is the 11-dimensional Planck mass.

We are interested in turning on a bulk field which corresponds to a fermion mass. The fermions in the theory living on the M2 branes transform in the {\bf 8}$'$ of the SO(8) R symmetry group. The operators $\lambda_i \lambda_j - \delta_{ij}\lambda^2/8$ are chiral and transforms in the ${\bf 35}_-$ of this group. Thus, their dimension does not change, and their coefficient, transforming also in the ${\bf 35}_-$ is a fermion mass \cite{aharony}.
Therefore, the bulk field which we have to turn on is a field strength $F_4^1$   oriented perpendicular to the brane, and transforming in the ${\bf 35}_-$ of SO(8) - which is anti self dual tensors. 

The 11d supergravity 4-form field strength satisfies the equation of motion:
$$d *_{11} F_4 = -{1 \over 2} F_4 \wedge F_4 \eqno(3)$$
The total field strength will contain both the background $F_4^0$ and the perturbation $F^1_4$. Thus $F_4=F_4^0+F_4^1$. We can reduce the 11 dimensional Hodge dual to an 8-dimensional one:
$$*_{11} F_4^1 = Z^{-1}(*F_4^1) \wedge dx^0 \wedge dx^1 \wedge dx^2 \eqno(4)$$
Combining (1)(3) and (4) we obtain the equation of motion for the perturbed field:
$$d[Z^{-1}(*F^1_4 - F^1_4)] = 0. \eqno(5)$$
Since $F_4^1$ can be written as the exterior derivative of a vector potential: $F_4^1 = d C_3^1$, the Bianchi identity is simply $d F^1_4 = 0$.

\subsection{Tensor Spherical Harmonics}
As we explained in the previous section, a fermion mass corresponds to an anti-self-dual 4-tensor SUGRA background. Other operators which may be of interest in this theory transform in ${\bf 35}_+$ and correspond to antisymmetric self-dual 4-tensors backgrounds. Thus
$$*T_{ijkl} \equiv {1 \over 4!} \epsilon_{ijkl}^{\ \ \ \ mnop}T_{mnop} = \pm T_{ijkl}, \eqno(6)$$
where the $+$ and $-$ are for ${\bf 35}_+$ respectively ${\bf 35}_-$.To make an (anti) self dual 4-tensor field on the space transverse to the M2 brane transforming in the same way, we can use $T_4$ or combine it with the radius vector to form
$$V_{mnpq}={x^r \over r^2}[x_m T_{rnpq} +x_n T_{mrpq} +x_p T_{mnrq} +x_q T_{mnpr}]. \eqno(7)$$
The forms $T_4$ and $V_4$ are normalized:
$$T_4 = {1 \over 4!} T_{mnpq} dx^m \wedge dx^n \wedge dx^p \wedge dx^q, \ \ \ \  V_4 = {1 \over 4!} V_{mnpq} dx^m \wedge dx^n \wedge dx^p \wedge dx^q. \eqno(8)$$ In addition we define
$$S_3 = {1 \over 3 !}  T_{mnpq} x^m \  dx^n \wedge dx^p \wedge dx^q.\eqno(8)$$
Since the general perturbation will be a combination of $T_4$ and $V_4$ multiplied by a power of $r$, the following relations  will be useful:
$$d S_3 = 4 T_4\ , \ \ \ d(\ln r) \wedge S_3 = V_4\ , \ \ \ d (r^p S_3) = r^p (4 T_4 + p V_4)\ ,$$
$$ d T_4 = 0\ , \ \ \ d V_4 = -4 d(\ln r) \wedge T_4\ , \ \ \ dr \wedge 
V_4 =0$$
$$*T_4 = \pm T_4\ , \ \ \ * V_4 = \pm (T_4 - V_4). \eqno (9) $$

In order to relate fermion masses to tensors it is convenient to group the 8 fermions and the 8 transverse coordinates in complex pairs:
$$z^1 = x^3 + i x^7\ ,\ \  z^2 = x^4 + i x^8\ ,\ \  z^3 = x^5 + i x^9\ ,\ \ z^4 = x^6 + i x^{10}\ ,\eqno(10a)$$
Similarly the fermions can be ``complexified'':
$$\Lambda^1 = \lambda^1 + i \lambda ^2\ ,\ \  \Lambda^2 = \lambda^3 + i \lambda^4\ ,\ \ \Lambda^3 =\lambda^5 + i\lambda^6\ ,\ \ \Lambda^4 = \lambda^7 + i \lambda^8 \ .\eqno(10b)$$
Under a rotation $z^i \rightarrow e^{i \phi_i} z^i$ the fermions transform as :
$$\Lambda^1 \rightarrow e^{i(-\phi_1+\phi_2+\phi_3+\phi_4)/2 } \Lambda^1$$
$$\Lambda^2 \rightarrow e^{i(\phi_1-\phi_2-\phi_3+\phi_4)/2 } \Lambda^2$$
$$\Lambda^3 \rightarrow e^{i(\phi_1+\phi_2-\phi_3+\phi_4)/2 } \Lambda^3$$
$$\Lambda^4 \rightarrow e^{i(\phi_1+\phi_2+\phi_3-\phi_4)/2 } \Lambda^4 \eqno(11)$$

If we give masses to the 4 complex fermions we preserve $\N=2$ supersymmetry    . Thus the Lagrangian will be perturbed with: 
$$\Delta L = {\rm Re} (m_1 \Lambda_1^2+ m_2 \Lambda_2^2+ m_3 \Lambda_3^2+ m_4 \Lambda_4^2)  \eqno(12) $$
This perturbation transforms under SO(8) like: 
$$ T =  {\rm Re} (m_1 d\bar z^1 \wedge dz^2 \wedge dz^3 \wedge d z^4+ 
m_2 d z^1 \wedge d\bar z^2 \wedge d z^3 \wedge d z^4+ $$
$$+ m_3 dz^1 \wedge d  z^2 \wedge d\bar z^3 \wedge d z^4+ 
m_4d z^1 \wedge dz^2 \wedge dz^3 \wedge d\bar z^4) \eqno (13)  $$
Regardless of the masses, this perturbation is invariant under the discrete $Z_2$ symmetry: 
$$z_1 \rightarrow i \bar z_1,\ \ z_2 \rightarrow i \bar z_2,\ \ z_3 \rightarrow i \bar z_3,\ \ z_4 \rightarrow - i \bar z_4.\eqno(14)$$ 
This symmetry has to do with the fact that our tensors are anti self dual. 
We will first be exploring the SO(4) symmetric configuration: $m^1=m^2=m^3=m^4 = m$. We can easily check that in this case only two components of $T$ will be nonzero: $ T_{3456} = - T_{789\ 10}= 4m$.  In chapter V we will be exploring generalizations to unequal masses.

\subsection{Linearized Perturbations}
As discussed in the previous subsection, the general form of the perturbation is 
$$F_4^1 = r^p(a T_4+b V_4) \eqno(15) $$
  We can use the Bianchi identity and (9) to simplify this to:
$$ F_4^1 = (a/4) r^p( 4 T_4+p V_4) = (a/4) d (r^p S_3).  $$
Using the equation of motion (5) and (9) we obtain after a few steps:
$$p^2+14p +24 \mp 24 = 0. \eqno (16)$$
We can see that for ${\bf 35}_-$ there are 2 solutions

$$p = -6\ ;\ \ \  F_4^1 \sim (R/r)^6 [4 T_4 - 6 V_4], \eqno(17a) $$
$$p = -8\ ;\ \ \  F_4^1 \sim  (R/r)^8 [4 T_4 - 8 V_4], \eqno(17b) $$

Translating to an inertial frame, and remembering that the $AdS_4$ radius $u=r^2$, we can see that the first perturbation is nonnormalizable, and corresponds to turning on a fermion mass in the gauge theory, while the second one is normalizable and corresponds to the vev of $\lambda \lambda$. These perturbations correspond to a field which is $AdS_4$ pseudoscalar and $S^7$ 3-tensor, and which satisfies an $S^7$ equation of ``self duality in odd dimension'' type \cite{ads4}.
The mass perturbation corresponds to:
 $$F_4^1 = \alpha (R/r)^6 [4 T_4 - 6 V_4] = d \left({\alpha (R/r)^6 S_3 }\right) , \eqno(18) $$
where $\alpha$ is the numerical constant that relates the boundary theory mass to the coefficient of the nonnormalizable bulk mode. From now on and throughout this paper we will be absorbing $\alpha$ into $m$.

\section {M5 Brane Probes}

In this section we consider a test M5 brane in the $AdS_4 \times S^7$ geometry, perturbed with $F_4^1$ flux. This is a relatively simple problem which contains most of the ``meat'' of the more complicated problem - that of the M2 branes becoming polarized into M5 branes. 

We will be first examining the case of 4 equal masses, which has an SO(4) symmetry between the 4 complex scalars. Our test M5 brane has the geometry $R^3 \times S^3$, and has M2 brane charge $n$. Thus 3 of its directions are parallel to the M2 branes which create the geometry, while the other 3 are wrapped on an $S^3$ inside $S^7$.

Turning on a M2 brane charge (parallel to that of the source M2 branes) on the M5 brane is done by turning on a 3 form field strength flux $F_3$ on $S^3$. This can be done, but it is not so straightforward. The 3 form field strength on the M5 brane is self dual - so naively turning on a flux on the $S^3$ cannot be done without turning on a flux in the other 3 directions. Moreover, we are in a background of $C_3$, so the gauge invariant object in the brane theory is not $F_3$ but $F_3-C_3$. Thus before proceeding with the computation we need to have a thorough understanding of the M5 brane action and of how self duality is achieved.

\subsection{The M5 Brane Action}

The action of an M5 brane is more complicated than that of D-branes because, as we said, the theory on the M5 brane contains a self dual 2 form field. We will only be interested in the bosonic part of this action. There are 2 approaches at writing an action for such a field. 

The first approach, by Pasti Sorokin and Tonin \cite{pst} consists in combining in a clever way the 3-form field with an auxiliary scalar field $a$ to form the action:
$$S_{PST} = -\int d^6 x \left[{\sqrt{- \det(g_{mn} + i \tilde H_{mn})} + \sqrt{-g} {1 \over 4 \p_r a \p^r a  } \p_m a (*H)^{mnp}H_{npq}\p^q a }\right]$$ 
$$ - \int \left[{C^6 + {1 \over 2} F \wedge C^3}\right] .\eqno(19) $$
Here $C^6$ and $C^3$ are the pullbacks of the M-theory forms, $F = dB$ is the field strength living on the brane, $H_{mnp} \equiv F_{mnp} - C_{mnp}$, $*$ represents the Hodge dual, and $\tilde H_{mn} = (*H)_{mnp} {\p^p a \over \sqrt{ \p_r a \p^r a }} $. 

The action has a Lorentz invariant form, and the self duality of $H$ is forced when integrating out the auxiliary field. The first term looks like a Born -Infeld term (and reduces to the normal Born - Infeld  term for a D4 brane),  the second term is a mixed term (which reduces to a part of the Wess-Zumino term of a D4 brane, but which unlike normal Wess-Zumino terms is not zero in the absence of background fields). The third term is a Wess-Zumino term. Using this approach the relative normalizations  of the 3 terms in the action, and the generalized formula  for background fields (19) can be easily found. Nevertheless, in order to compute anything using the first 2 terms one has to fix some of the gauge symmetries. 

The second approach, by Perry and Schwarz \cite{ps} consists in picking a special direction, and thus maintaining only  5-d explicit Lorenz invariance (although the theory secretly is 6-d Lorentz invariant). The 6d metric $G_{\hat \mu \hat \nu}$ contains 5d pieces $G_{\mu \nu}, G_{\mu 5}$ and $G_{55}$. Hatted indices denote 6 dimensional quantities, and unhatted ones represent 5 dimensional ones. The self dual antisymmetric tensor is represented by a $5 \times 5$ antisymmetric tensor $B_{\mu\nu}$, and its curl $F_{\mu\nu\rho} = \p_{\mu} B_{\nu\rho } + \p_{\nu} B_{\rho \mu }+ \p_{\rho} B_{\mu\nu } $.
The action obtained is 
$$S_{PS} = -\int d^6 x (L_1+L_2+L_3) ,$$
where 
$$L_1 = \sqrt{-\det\left({G_{\hat \mu \hat \nu} +{i  G_{\hat \mu \rho} G_{ \hat \nu \lambda} \tilde H^{\rho \lambda} \over \sqrt{-G_5}} }\right)} $$
$$L_2 = {1\over 4}  \tilde H^{\mu\nu} \p_5 B_{\mu\nu},  $$
$$ L_3 =- {1\over 8}  \epsilon_{\mu\nu\rho\sigma\lambda} {G^{5 \rho} \over G^{55}  } \tilde H^{\mu\nu} \tilde H^{\sigma\lambda},\eqno(20) $$
where $ \tilde H^{\mu\nu}  = {1 \over 3 !} \epsilon^{\mu\nu\rho\sigma\lambda}   F_{\rho\sigma\lambda}   $, and $G_5 = \det G_{\mu\nu}$, and the $\epsilon$ symbol is purely numerical.

This action can be used directly for explicit computations, and can be obtained  from the PST action with no external background upon fixing $\p_{\mu} a = \delta_{\mu}^5$, and $B_{\mu 5} = 0$ . $L_1$ is obtained from the Born-Infeld term, and $L_2+L_3$ are obtained from the mixed term.
Self duality (which in this approach appears as an equation of motion) in the limit of a free theory in the gauge $B_{\mu 5} = 0$  is $  \tilde H^{\mu\nu} = \p_5 B_{\mu\nu} $. Note that $\tilde H^{\mu\nu}$ is not a tensor, since $\epsilon$ is numeric 

If the theory is interacting, the self duality relation is more complicated, and can be found in its full splendor in \cite{schwarz}. For the cases we are interested in, where only 345 and 012 fields are turned on, the equations simplify to give:

$$   \p_5 B_{\mu\nu}  = K_{\mu\nu} = {\sqrt{-G}G_{\mu\mu'}G_{\nu\nu'} \tilde H^{\mu'\nu'} \over -G_5 \sqrt{1 + H_{\mu\nu\rho} H^{\mu\nu\rho} } }.
\eqno(21)  $$ 
In the language of \cite{schwarz}, if only 345 and 012 fields are turned on,  $z_1^2 =2 z_2$, which makes the formulas simplify. Under this assumption, also 
$$L_1 = \sqrt{-G} \sqrt{1 + H_{\mu\nu\rho} H^{\mu\nu\rho} }. $$

In \cite{schwarz} the action is given for a general gravitational background, but not for a background with a 3-form field turned on. The PST action however is given in the presence of a background 3-form field, but it is hard to use for explicit computations. Fortunately, we know \cite{costin,pst} how to obtain the PS action from the PST action without a background field. Therefore, we expect to obtain a generalization of the PS action by gauge fixing the PST action with background field. 
This can be easily done, and the only change in the PS action is : $ \p_5 B_{\mu\nu}  \rightarrow  \p_5 B_{\mu\nu}  - C_{5 \mu\nu}$, $ \tilde H^{\mu\nu}  \rightarrow {1 \over 3 !} \epsilon^{\mu\nu\rho\sigma\lambda} ( F_ {\rho\sigma\lambda} -  C_ {\rho\sigma\lambda} )$.

\subsection{A Toy Problem with the M5 brane action}

In this section we will do a toy problem in which the interplay of the 2 formulations of the M5 brane action is shown. Let us consider a flat M5 brane extended in the 012345 directions in flat 11d,  in a background of $C_{012}$ and $C_{345}$. We turn on  a nonzero $F_{345} $, and we select ``2'' as our special direction in the PS action. We call $f = F_{345}-C_{345}$. Thus $\tilde H^{01}=f$, and the Born Infeld term is $L_{BI} = - \sqrt{1+f^2}$. The equation of motion (21) gives:
$$\p_2 B_{01}-C_{201} = {f \over \sqrt{1+f^2}}, \eqno(22)$$
and thus 
$$L_2 = - {f^2 \over 2 \sqrt{1+f^2}}.$$
Since the metric is diagonal, $L_3=0$. We can read off the Wess-Zumino term from the PST action. In the gauge we chose, $F_{012} = \p_2 B_{01}$ is given by (22). Note that the action goes like $f^2$ for small $f$ and like $f$ for large $f$. This is also characteristic to the D-brane action. Note that applying naively the weak coupling version of the self-duality would give us an action growing like $f^2$ for large $f$, which is nonphysical (it does not reduce to Born Infeld).

Let us try to get some intuition about the physics of the problem. For small $f$ and no background fields, turning on a flux in the 345 direction induces the turning on of a flux in the 012 of the same magnitude (22). What this tells us is that if we dissolve an M2 brane in an M5 brane their fields force (by the SUGRA equations of motion) the appearance of a field which couples with an orthogonal M2 brane.
However if $f$ is very large, (22) tells us that the dual M2 brane charge asymptotes to 1.

\subsection{The M5 brane probe}

We consider a large number of M2 brane along the $012$ directions and an M5 brane with 3 directions wrapped on an $S^3$. Since we know the effect of rotations in the 3-7, 4-8, 5-9, and respectively 7-10 planes on all the fields,  we can assume the plane of the sphere to be $3456$. Let us denote by $\hat \epsilon_{ijkl}$ the numerical antisymmetric tensor restricted to the $3456$ plane. We also give the M5 brane an M2 brane charge $n$, by turning on a 3-form field strength along $S^3$:
$$F_3 = {4 \pi n \over M_{11}^3r^4 3!} \hat \epsilon_{ijkl} x^i dx^j \wedge  dx^k  \wedge dx^l. \eqno(23) $$
We assume $n < N$, so the effect of the M2 brane charge of the probe on the background can be ignored. From (12) (13) and (18) we find that $T_{3456}= 3m $, and thus 
$$C_3^1= \left({R \over r}\right)^6 S_3 = 3m   \left({R \over r}\right)^6  {\hat \epsilon_{ijkl} \over 3!} x^i dx^j \wedge  dx^k  \wedge dx^l. \eqno(24) $$
For further reference we can also express $C^1_{\theta \phi \alpha}$ and $F_{\theta \phi \alpha }$ using the angles of the 3 sphere, by noticing that $\hat \epsilon_{ijkl} x^i dx^j \wedge  dx^k  \wedge dx^l |_{S^3} = 3!\ r^4 \sin^2 \theta \sin \phi \ d \theta \wedge d \phi \wedge d \alpha $.  $C_3^0$ is given by (1).
The M theory 6 form is the dual of the 3 form and can be found using
$$d C_6 - {1\over 2} C_3 \wedge F_4 = d F_7 = d * F_4. \eqno(25)$$
Using (4), (17) and the relations in (9), we obtain 
$$dC_6 =[ Z^{-1}(*F_4^1-F^1_4)  + {1 \over 2}d( Z^{-1} C^1)]  \wedge  dx^0  \wedge dx^1 \wedge dx^2 = 0. \eqno(26)$$
Thus the first term in the Wess-Zumino action gives no contribution. This is different from \cite{joe}, where the nonzero 6-form background gave one of the leading contributions to the action. 

Since we have spherical symmetry, the value of the action will be the same at every point on the 3-sphere. To make the computation more explicit we concentrate on the point $x^6 = r$, and we chose ``2'' as our special direction. Thus:
$$H_{345} = H_{\perp} = - \left[{{4 \pi n \over M_{11}^3 r^3} - 3mR {R^5 \over r^5}  }\right] = - {A \over r^3}  - C_{345},\eqno(27) $$
where $A \equiv 4 \pi n /M_{11}^3$ 
We are interested in the limit when the M2 brane charge of the M5 brane is bigger than its M5 brane charge. This means $ n >> \sqrt{N} $. Therefore we expect the first term in $H_{345}$ to be dominant. We separate the 6 dimensional metric into perpendicular and parallel parts and denote by $ G_{\perp }$ and $G_{\parallel}$ their respective determinants. Using the equation of motion (21) we obtain 
$$\p_2 B_{01} - C_{201}=  -H_{\perp} G_{\perp}^{-1}  {\sqrt{- G_{\perp} G_{\parallel }} \over \sqrt{1 + H_{\perp} ^2G_{\perp}^{-1} }}
= - H_{345} G^{33}G^{44}G^{55}{\sqrt{-\det G_{\hat \mu \hat \nu}} \over  \sqrt{1+  H_{345} H^{345}  } } \eqno(28)$$
Since $C_{012}$ is known, (28) gives us the value of $F_{012}$, which couples with $C_{345}$ in the Wess Zumino term. For large M2 brane charge , $-C_{201}$ is very close to the right hand side of (28), and basically the Wess-Zumino term containing $F_{012}$ is negligible. 

This is very interesting, and definitely not a coincidence. What we discovered is that the equations of motion of an M5 brane with relatively large M2 brane charge in a geometry given by (1), give rise to a ``dual'' 3 form equal to the background 3-form field of this geometry, for any M2 brane charge which is large enough. Note that this is independent of $Z$, and even of the shape of the M5 brane (the $G_{\perp}$'s cancel out). Thus the M5 brane Lagrangian somehow knows about the M-theory equations of motion. This is an interesting, if somewhat not expected connection which deserves further study.

We have all the pieces needed to compute the full potential for an M5 brane in this geometry. We also integrate over the sphere, which will give us a potential energy per unit length. The relevant parts of the potential are:

$${-S_{BI} \over 2 \pi^2 V}  = r^3 \sqrt{{1 \over Z}\left[{ 1 + {H_{345}^2 \over  Z} }\right] } \approx  {A \over Z} +   { r^6  \over 2 A } + { r^3 C_{345} \over  Z} .\eqno(29a)$$
$${-S_{mixed} \over 2 \pi^2 V}  =  {-r^3 \over 2} { H_{345} H^{345} \sqrt{-G} \over  \sqrt{1+  H_{345} H^{345} } } \approx -{A  \over 2  Z} + { r^6  \over 4 A } - { r^3 C_{345} \over 2 Z}   . \eqno(29b)$$
$${-S_{WZ}\over 2 \pi^2 V}   = -{r^3 \over2} [- 2 C^6_{012345}+ C_{012}F_{345} - F_{012}C_{345}] \approx - {A \over 2 Z},  \eqno(29c)$$
where the approximation is in the large $n$ limit. We included the $C^6$ component which in this case is 0, in order to see how it combines with $C_{345}$ in this action.

As expected, the dominant contributions of the 3 actions come from the M2 brane charge of the M5 brane, and they cancel. The second terms in (29a) and (29b) represent the gravitational energy of the M5 brane. In the absence of a background $C_{3}$ they would cause the M5 brane to collapse on the stack of M2 branes.

Since we only worked at order $m$ in perturbation theory, there can be another term proportional to $m^2 A$ which has the same relevance as the first 2. Indeed, we expect an order $m^2$ correction in $C_{012}$, via (3). Since $C_{012}$ couples with $F_{345}$, we can see that this correction will be relevant. It can also be easily seen that the contribution  of this term to the potential goes like $r^2$.

Thus, the dominant part of the potential is:
$${-S \over 2 \pi^2 V} = {3 r^6 \over 4 A} - {4m \over 2} r^4  + c A m^2 r^2, \eqno(30)  $$
where $c$ is not yet determined. Since we have 4 supercharges, $c$ can be computed easily. Indeed, the potential is the square of the derivative of the superpotential, so $c$ is obtained by simply completing the square. Before proceeding with this we need to write the potential in a more general form. As we mentioned at the beginning of this chapter, we restricted our attention the 3456 plane. The general SO(4) invariant brane configuration is obtained by rotating the sphere of radius $r$ in the 3-7, 4-8, ... planes by the same angle $\theta$. This configuration is parametrized by $z = r e^{i \theta}$. Examining the effect of rotation on the terms we had, we see that the action  will be:
$${-S\over 2 \pi^2 V} = {3 |z|^6 \over 4 A} - {2m} {\rm Re}(z^3 \bar z) + {4|z|^2 m^2  A \over 3} = {3 \over 4 A}|z^3+4 z m A /3|^2,  \eqno(31)$$
where the last term was obtained by completing the square.
This has a supersymmetric minimum in the 3456 plane at
$${r^2} = {4 m A \over 3}. \eqno(32)$$

There is however an extra case to consider. Our mass perturbation (13) is invariant under (14), which flips the M5 brane from the 3456 plane to the 879 10 plane. Thus, there will also exist a supersymmetric minimum corresponding to an anti M5 brane of the same radius, in the 789 10 plane.   We will have more comments on the superpotential which generates the action (31) in chapter V.

This is what we advertised: a test M5 brane in the background formed by M2 branes with $F_4$ flux has a ground state at nonzero $r$.
As we mentioned, the equivalent string theory picture \cite{joe} can be interpreted in 2 ways - as D3 brane polarization or as a ground state for a D5 brane with D3 brane charge. Unfortunately, no one has studied the polarizability of M2 branes, so we can only assume it takes place by analogy with the string theory case.

\section{The Full Problem - Warped Geometry }

As we have seen in the previous chapter, turning on fermion masses polarizes the M2 brane. We now consider the case of the $N$ M2 branes distributed uniformly on one or several 3-spheres, with M5 brane charges. When the 5 brane charges are relatively small, the background geometry will still be given by (1), but $Z$ will be different. Since we will lose most of the symmetry, it appears that the problem will be far harder than the one with a probe M5 brane. Fortunately, like in \cite{joe}, the action does not change. so there is no more work to do. However, here the ``conspiracy'' which makes this work is far more unexpected.

\subsection{The Geometry}

As known from time immemorial, the geometry created by a distribution of M2 branes is still given by (2), but with $Z$ being the superposition of the harmonic functions sourced by each brane.
If for example the M2 branes are spread over a 3-sphere of radius $r_0$ in the  $3456$ plane, the new Z-factor will be
$$Z = {2 \over \pi} \int_0^{\pi}{R^6 \sin^2 \theta \  d \theta \over (r_2^2+r_1^2+r_0^2-2 r_0 r_1 \cos \theta)^3 }  = {R^6 \over (r_2^2+(r_1+r_0)^2)^{3 \over 2}  (r_2^2+(r_1-r_0)^2)^{3 \over 2} }, \eqno(33)$$ 
where $r_1$ and $r_2$ are the radii in the $3456$ and respectively $789\ 10$ planes. When the M2 branes are distributed over several such spheres, in the $3456$ and $789\ 10$ planes, Z will be a sum of such terms, properly normalized.

The field equation for an antisymmetric antiselfdual perturbation is again:
$$d[Z^{-1}(*F^1_4 - F^1_4)] = 0. \eqno(34)$$
and the Bianchi identity is $d F_4^1 = 0$. The behavior of the solution at infinity is given by the boundary theory, and is the same as for trivial $Z$. There is also a magnetic source corresponding to the 5 brane, but this creates a normalizable mode, which is subleading at $\infty$.
We can perform the same clever trick as in \cite{joe}. From (34) we can derive:
$$d* [Z^{-1}(*F_4^1-F_4^1)] = 0. \eqno(35)$$
Therefore, by the Hodge decomposition we derive that $ [Z^{-1}(*F_4^1-F_4^1)]$  is harmonic, and thus equal to its value at $\infty$: 
$$  Z^{-1}(*F_4^1-F_4^1) = -2 T_4. \eqno(36)$$
In particular, $C_3^1$ and $C_6$ will change, but (26) implies that the combination $ [C_6  + {1 \over 2} C_3^1 \wedge C_3^0] $ will not change.  This same combination appears in the M5 brane action (29), namely $C_{012345} - {1\over 2Z} C_{345}$. We are definitely seeing a ``conspiracy'' - the factor of $-1/2$ came from both the Born-Infeld and the mixed term of the M5 brane Lagrangian, and provides exactly the combination which is unchanged when we change $Z$. 

There is one more thing to consider, the effect of the M5 brane charge on itself. The M5 brane is a magnetic source for the 3-form field, and it appears as a source in the right hand side of the Bianchi identity. Nevertheless, the M5 brane only couples with the combination (36), which remains unchanged, and thus it is unaffected by itself.

\subsection{The solutions}

Let us consider first the potential felt by a probe M5 brane with M2 brane charge in the geometry created by several shells of M2 branes. This is still given by (29) but with a different $Z$. The leading contributions cancel as usually. The first 2 terms in (30) will not change, and by supersymmetry, the third term will not change either. Therefore, the potential will be independent of the distribution of the sources.

We would like now to consider the potential felt by the full set of M2 branes. This can be found by bringing the branes one by one from $\infty$. In our case, like in  \cite{joe} the potential felt by a brane does not depend on the distribution of the others, so as explained there, the potential is the same as in the probe case.

If the 4 masses are equal, a general ground state is a configuration consisting of M2 brane 3-sphere shells with charges $n_b$ in one of the planes $3456$ and $789\ 10$. The potential will be:
$$ {-S \over 2 \pi^2 V} = \sum_b{{3 \over 4 A_b}\left|{z_b^3 - {4 Q_b z_b m A_b \over 3}}\right|^2 }, \eqno(37)$$
where $Q_b$ is by convention 1 for M5 and -1 for anti M5 branes, and  $A_b \equiv 4 \pi n_b / M_{11}^3$. There is however one more condition which we ignored. In order for the geometry to be valid, the M5 brane charge density of the shells should be smaller than their M2 brane charge. This means:
$$n_b >> \sqrt N \eqno(38) $$
Thus the possible configurations will be given by distributions satisfying (38). 

We can see that there is a large number of discrete vacua, corresponding to combinations of charges $n_b$, adding to $N$ and satisfying (38), in both the 3456 and the 789 10 planes. It is a straightforward exercise to compute the normalizable modes created by the M5 branes, in each of these vacua. The coefficient of these normalizable modes gives the value of a condensate which contains the fermion condensate and its supersymmetric partners. The vacua will be distinguished by the values of these condensates. The $Z_2$ symmetry will relate the vacua with M5 branes replaced by anti M5 branes ($Q_b \rightarrow -Q_b$).

Unfortunately, we cannot interpret the vacua in any way as corresponding to broken gauge symmetries some of which are restored when the M5 branes are coincident, since the theory living on $N$ M2 branes has mysterious degrees of freedom.
Just for fun we may observe that a relation similar to (82,83) in \cite{joe} also holds here. Namely, if $n_1 \times n_2 = N$, descriptions with $n_1$ coincident M5 branes and with $n_2$ coincident anti M5 branes have complementary ranges of validity. A speculative mind may see in this a sign of some duality, but we will refrain from further commenting on that.

\section{Unequal masses}

We can try to generalize the previous construction for the case of unequal masses. Since we will only be interested in the limits when one or two masses go to 0, and since want to keep the presentation simple, we will keep  $m_3=m_4\equiv m$ and vary $m_1$ and $m_2$. Unlike the previous case, where the general configurations was SO(4) invariant, here we only have SO(2) symmetry, so we expect the M2 branes to become polarized into an ellipsoid with 2 equal axes. 
Again we can restrict to the 3456 plane and then obtain more general configurations by phase rotations. The ellipsoid will be parametrized:
$$x^3 = a \ r \cos\theta $$
$$x^4 = b \ r \sin\theta\cos\phi $$
$$x^5 = r \sin\theta\sin\phi\cos\alpha $$
$$x^6 = r \sin\theta\sin\phi\sin\alpha  \eqno(39)$$
From (23) and (24), we see that on the ellipsoid  $C^1_{\theta \phi \alpha}$ is multiplied by $ab$ and $F_{\theta \phi \alpha }$ is unchanged. Also in the $\theta \phi \alpha $ coordinates
$$G_{\perp} = r^6 Z  \sin^4 \theta \sin^2 \phi (a^2 \sin^2 \theta \cos^2\phi + a^2 b^2 \sin^2\theta \sin^2\phi+b^2 \cos^2 \theta) \eqno(40)$$
We noted that for any distribution of brane, the M5 brane equations of motion create a dual field almost equal to the background $-C_{012}$, so there will be no new contribution from the Wess-Zumino term. 
The dominant terms of the potential do not depend on $G_{\perp}$, so they will have the same value as before, and they will cancel as usually.
The terms which before were proportional to $ r^6 / A$ will now be variable on the ellipsoid, and their value will be:
$$V_{\sim r^6} \sim \int_{E^3}{\sqrt{G_{\parallel}} {G_{\perp} \over H_{\theta \phi \alpha }}} \eqno(41)$$
What one might have been afraid of (the integrand proportional to $ \sqrt{G_{\perp}} $ - which would have made the integral elliptic) does not happen. The integral can be easily worked out to give:
$${-S_{\sim r^6} \over 2 \pi^2 V} = {3 r^6 \over 4 A}{ a^2 + b^2 + 2 a^2 b^2  \over 4} \eqno(42)$$
The term proportional with $C_3$ will be multiplied by $ab$, and will change also because of (12).
Putting back the phases corresponding to rotations, and remembering that  $|z_3|=|z_4|=r$, $|z_1| = a\ r$, and $|z_2|= b\ r$ we obtain the potential to be:

$$ {-S \over 2 \pi^2 V} = {3 \over 16 A} \left({ 2 |z|^2 |z_1|^2 |z_2|^2 + |z|^4 |z_1|^2+ |z|^4 |z_2|^2 }\right) - {1\over 2} {\rm Re}(2m z_1z_2 z\bar z+m_2 z_1 \bar z_2 z z+m_1 \bar z_1 z_2 z z) $$ 
$$+ {A \over 3}(2 m^2 |z|^2+m_1^2 |z_1|^2+ m_2^2 |z_2|^2)  . \eqno(43) $$
where the second line was added to complete the square, as required by supersymmetry. We can illustrate better the SUSY nature of this action by writing it as:
$$ {-S\over 2 \pi^2 V} =  {3 \over 16 A} (2 | z z_1 z_2 - 4 m A z /3|^2 +  | z^2 z_2 - 4 m_1 A z_1 /3|^2 +| z^2 z_1 - 4 m_2 A z_2 /3|^2  ) \eqno(44)$$
This potential has a supersymmetric minimum at:
$$z_1^2 = {4 A \over 3} \sqrt{m^2 m_2 \over m_1} \ \ \ \  
z_2^2 = {4 A \over 3} \sqrt{m^2 m_1 \over m_2} \ \ \ \  
z^2 = {4 A \over 3} \sqrt{m_1 m_2 m\over m} \eqno(45)$$
Thus, the branes will polarize into an ellipsoid in the 3456, respectively 879 10 planes with the axes given above.

We will now try to write a few comments about the superpotential corresponding to the action(44). We can see that if we assume our fields to be the complex scalars $z_i$, we can write a superpotential of the form
 $$ W \sim z_1 z_2 z_3 z_4 - {2A \over 3}\sum_i{m_i z_i^2} \eqno(46)$$
In the case of polarized D-branes, a similar superpotential came from a
superpotential originally of the form:

$$W \sim {\rm tr} (\Phi_1 [\Phi_2,\Phi_3]), \eqno(47) $$
perturbed with a mass term. The $\Phi_i$ were the $N \times N$ scalar fields on the brane, corresponding to position in spacetime. The mass terms forced the ground state $\Phi_i$ to become noncommutative, which corresponded to polarization of the Dp brane into a D(p+2) brane.
 
In our case however we do not know what form the scalars representing the position of the M2 brane have (they are definitely not $N \times N$ matrices). Therefore, we do not know what form the equivalent of (46) will have. What we do know is that it will involve all the 4 scalar fields, and that adding mass terms will cause the ground state fields to become ``noncommutative'' (more rigorously speaking - to modify the ground state fields so that the term which contains all the 4 fields will be nonzero). This can give an intuitive picture for the polarization of the M2 brane into a brane with 3 extra dimensions.

We also observe that the superpotential (46) has a ``classical'' form. More precisely, it looks like the superpotential of a theory with 4 massive hypermultiplets, and does not contain nonperturbative terms. Since we are at very strong coupling  and we do not know the degrees of freedom of the theory, we can only suggest that this happens because of $N$ being large. 

We can also make a few comments about cases when some of the masses go to 0.  Using (45) we can see that if we take one mass to 0, the ellipsoid degenerates into a line of very long length, which corresponds to the theory having a moduli space. Intuitively we can see that if $z_1$ has no mass,  it is a modulus. Nevertheless, this moduli space is not protected by supersymmetry, and can in principle be lifted by corrections. If we take $m_1, m_2 \rightarrow 0$ keeping $m_3=m_4=m$, we restore $\N=4$. The ellipsoid degenerates into a pancake of radius $r^2=\sqrt{4A m^2/3}$. It would be interesting to give an interpretation for this in the framework of theories with 8 supercharges.

\section{More about the theory on the brane}

Since not too much is known about the $\N=2$ theory whose dual we constructed, we can only use the duality one way: to interpret the possible M-theory configurations from the point of view of the $\N=2$ theory.

\subsection{Domain Walls}

Since our theory has multiple vacua, they can be separated by domain walls. Let us consider the domain wall between the vacuum corresponding to all the 2-branes polarized into one M5 brane, and the vacuum with 2 M5 branes of charges $n_1$ and $n_2$. Since the first 3-sphere has radius $r \sim \sqrt N$, and the concentric 3-spheres will have radii $r_i \sim \sqrt{n_i}$, they will both bend and meet at an intermediate radius $r_0$. By charge conservation, another M5 brane should come out of the junction. Therefore, the domain wall should correspond to a M5 brane filling the 3456 ball of radius $r_0$ and extended in the 01 directions. Since the M5 brane has 2 longitudinal and 4 transverse directions, it feels no warp factor, and its tension will be 
$$\tau_{1} = {M_{11}^6 \over (2\pi)^5} {\pi^2 r_0^4 \over 2} \sim m^2 N^2. \eqno (48)$$
The tension will have another piece, $\tau_2$, which comes from the bending of the branes.

In the case of vacua with the same number of M5 branes, the branes will just bend into each other, and there will be no object required by charge conservation to fill the 4-ball. Thus the tension will only have one piece coming from the bending of the branes. It is not hard to compute $\tau_2$, although we have not done it here. 

We can compare $\tau_1$ with the tension of a supersymmetric domain wall, which is given by the difference of the superpotentials in the 2 phases. Using (46) and (37) we can see that for two generic vacua
$$ \tau_{DW} \sim |\Delta W| \sim m^2 N^2. \eqno(49)$$
This has the same dependence on $m$ and $N$ as $\tau_1$. It will be an interesting exercise to show that the construction with bent M5 branes reproduces the superpotential calculation for supersymmetric domain walls. Even if the exact normalization of the superpotential is not known, the matching of the dependence of the tension on the different $n_i$'s would be a beautiful result.  

Since naively the superpotentials are the same in 2 vacua related by the $Z_2$ symmetry, it may appear from (49) that the tension of the domain wall between them is 0. Similarily, (49) implies that the tension between vacua characterized by M2 brane charges $n_i$ and $n_i'$ is zero, if $ \sum n_i = \sum n_i' =N$ and  $ \sum n_i^2 = \sum n_i'^2$ . This appears to contradict the expectation that the bending tension $\tau_2 \neq 0$. Nevertheless, since we do not know the relative sign of the superpotentials in the 2 phases, and we also do not know if the domain walls are supersymmetric, there is no contradiction.

\subsection{Condensates, Instantons, etc.}

The only candidate for an instanton (an object with spacetime dimension 0) is an M2 brane wrapped on the $S^3$. Nevertheless, this configuration is unstable. The wrapped M2 brane can attach to the M5 brane and slide off.

As we mentioned, the coefficient of the normalizable mode created by the brane configurations gives the value of the condensate which contains the fermion bilinear and its supersymmetric partners. This can be straightforwardly computed, although it is not done here. The other objects which may exist in this theory also need a more thorough investigation.

\section{Conclusions and future directions}
Most of the conclusions of this construction are identical to the conclusions of \cite{joe}, and probably the main one is that M theory  resolves the naked singularity on might have expected to obtain by turning on fermion masses.

Lots of things remain to be done and understood. The tension of the domain walls, as well as their shape are well within our reach. A bit hard is to get an idea about what do the plethora of vacua of the theory represent. 

Another interesting thing which we observed was the interplay of the 3 pieces of the M5 brane Lagrangian, and how the turning on of a large flux along the $S^3$ induced via the M5-brane equations of motion a dual field equal to the background 3-form. This is an interesting connection between the equations of motion of the M5-brane and of 11d supergravity which deserves further study. We should also mention that this is to our knowledge the first time when the M5 brane bosonic actions (both by Pasti, Sorokin and Tonin, and by Perry and Schwarz) were used in a direct calculation. The cancellation of main contributions in the potential (as required by supersymmetry), and the fact that the subleading terms reproduce a supersymmetric potential are nontrivial consistency checks for these actions.

The case when one or two hypermultiplet masses are brought to 0 also awaits a more thorough investigation. More can be said about the moduli of these theories, and possible connections can be made with the vast literature on theories with 8 supercharges.

There is also a relatively well developed subject dealing with generalizations of the $AdS_4 \times S^7$ duality to different less supersymmetric versions corresponding to various 7-manifolds. In particular the perturbation which preserves $G_2$ symmetry (and which is a combination of the self-dual and anti self dual field strengths) has been identified \cite{rey} as giving rise to the flow from the $\N=8,\  AdS_4 \times S^7$ vacuum to the $G_2$ invariant Englert vacuum. The tools developed in this paper can be used to learn more about that flow.

The theory whose M-theory dual we found still remains very mysterious. Nevertheless we were able using this construction to understand quite a few things about it. It is the (probably overoptimistic) hope of the author that this approach may bring us closer to a more complete understanding of this theory and of the M-theory degrees of freedom.

{\bf ACKNOWLEGEMENTS}: I'm deeply indebted to Joe Polchinski for his help and guidance through this project, and to Costin Popescu for his help in understanding the M5 brane action. I also profited from very stimulating conversations with Duiliu-Emanuel Diaconescu, Alex Buchel, Aleksey Nudelman, Andrew Frey, Mitesh Patel, and Simion Hellerman. This work was supported in part by NSF grant PHY97-22022.

\end{document}